\documentclass[12pt,preprint]{aastex}

\newcommand{\be}{\begin{equation}}
\newcommand{\ee}{\end{equation}}
\newcommand{\beq}{\begin{eqnarray}}
\newcommand{\eeq}{\end{eqnarray}}

\shorttitle{The Magnetic Structure of \it{TRACE} loops}
\shortauthors{L\'opez Fuentes \& Klimchuk}

\received{2005 July 14}
\begin{document}

\title{The Magnetic Structure of Coronal Loops Observed by {\it TRACE}}  

\date{}

\author{M. C. L\'opez Fuentes\altaffilmark{1}, J. A. Klimchuk} 
\affil{Naval Research Laboratory, Code 7675, Washington, DC 20375}

\author{P. D\'emoulin}
\affil{Observatoire de Paris, LESIA, F-92195 Meudon Principal 
Cedex, France}

\altaffiltext{1}{also at School of Computational Sciences, George Mason 
University, 4400 University Drive, Fairfax, VA 22030}

\begin{abstract}
Previous studies have found that coronal loops have a nearly uniform thickness, 
which seems to disagree with the characteristic expansion of active region 
magnetic fields.  This is one of the most intriguing enigmas in solar physics.  
We here report on the first comprehensive one-to-one comparison of observed loops 
with corresponding magnetic flux tubes obtained from cotemporal magnetic field 
extrapolation models.  We 
use EUV images from {\it TRACE}, magnetograms from the MDI instrument on 
{\it SOHO}, and linear force-free field extrapolations satisfying 
$\nabla \times \bf{B} = \alpha \bf{B}$ with $\alpha$ equal to a constant.  
For each loop, we find the particular 
value of $\alpha$ that best matches the observed loop axis and then construct 
flux tubes using different assumed cross sections at one footpoint (circle and 
ellipses with different orientations).  We find that the flux tubes expand with 
height by typically twice as much as the corresponding loops.  We also find that 
many flux tubes are much wider at one footpoint than the other, whereas the 
corresponding loops are far more symmetric.  It is clear that the actual coronal 
magnetic field 
is more complex than the models we have considered.  We suggest that the 
observed symmetry of loops is related to the tangling of elemental 
magnetic flux strands produced by 
photospheric convection.  
\end{abstract}



\section{Introduction}
\label{intro}


Soft X-ray and EUV observations have revealed that the solar corona is a 
very hot and highly structured medium (e.g., Orrall 1981; Bray et al. 1991). 
It is clear that the magnetic field plays a dominant role in structuring the plasma, and it is very likely that the field also plays a fundamental role in the heating.  Understanding the detailed properties of the coronal magnetic field is therefore a high priority.  One particular property---the apparent uniformity in the thickness of coronal loops and their associated magnetic flux tubes---is very puzzling and could be a vital clue as to the origin of coronal heating.

Because the magnetic Reynolds number is so large in the corona, the plasma and magnetic field are ``frozen" together.  Observed structures such as soft X-ray and EUV loops trace out magnetic field lines that are rooted in the photosphere.  Throughout much of the corona, and especially in the coronal part of active regions, the magnetic pressure dominates the plasma pressure ($\beta << 1$), and the field is approximately force-free such that
\be
\nabla \times \bf{B} = \alpha \bf{B} , \label{eq:ff}
\ee
where $\alpha$ is generally field-line dependent.
Under these conditions the strength of the field must tend to decrease away from the solar surface, and a majority of the flux tubes that make up the field must therefore expand with height.  This is not required of all flux tubes, however.  The overall expansion of large-scale magnetic configurations such as active regions is well established, but thin plasma loops are observed to have a nearly constant thickness along their length (Klimchuk et al. 1992).


This surprising result has been confirmed by Klimchuk (2000) and Watko \& Klimchuk (2000) who studied sizable collections of loops observed by the Soft X-ray Telescope (SXT) on {\it Yohkoh} and the {\it Transition Region and Coronal Explorer (TRACE)}, respectively.  The loops are only slightly wider at their midpoints than at their footpoints, implying a much smaller expansion than is present in standard magnetic field models.  In addition, there is only modest variation of width along each loop, suggesting that the cross-section must be approximately circular if the field has non-zero twist.  


Klimchuk, Antiochos \& Norton (2000) suggested that the X-ray and EUV loops might correspond to significantly twisted magnetic flux tubes that are surrounded by relatively untwisted field and faint plasma.  Parker (1977) and others had shown that the magnetic tension associated with the azimuthal component of the field would cause a constriction in the cores of straight twisted tubes.  Klimchuk et al. argued that this constriction would be greater in the thicker, i.e., higher, parts of realistic curved loops.  They constructed three-dimensional force-free field models which showed that twist can indeed promote thickness uniformity, but probably not to the degree implied by observations.  The models also indicate that twist tends to circularize the loop cross section in the corona.


One important limitation of the observational studies cited above is that they did not compare the {\it Yohkoh} and {\it TRACE} loops with the corresponding magnetic flux tubes obtained, for example, from magnetic field extrapolation models based on magnetograms observed at the same time.  Instead, the studies compared the typical expansion of observed loops with the characteristic expansion of generic magnetic field models.  Since not all of the flux tubes in a realistic field configuration are expected to expand, it is possible that the observed loops correspond to the subset of non-expanding tubes.  If that were the case, it would have important physical implications.  
For example, Longcope (1996) has suggested that coronal heating comes 
from reconnection at magnetic separators and that the magnetic field near separators has minimal expansion.  The work we present here is the first 
comprehensive study involving one-to-one comparisons of observed loops with their corresponding flux tubes. 

Our strategy is to compute linear force-free extrapolation models based on photospheric magnetograms from the Michelson Doppler Imager (MDI) on the {\it Solar and Heliospheric Observatory (SOHO)} and to compare them with carefully coaligned and nearly cotemporal images from {\it TRACE}.  We vary the force-free parameter $\alpha$ in equation (\ref{eq:ff}) to obtain the best possible fit between the model field and the observed loop, and we construct flux tubes by tracing field lines assuming a variety of possible cross sections at the footpoint.  We then compare the expansion of the loops and corresponding flux tubes and evaluate the extent to which they agree.

The paper is organized as follows. 
Section~\ref{loops} concerns the plasma loops observed by {\it TRACE}, henceforth referred to simply as ``loops." 
Section~\ref{fluxtubes} concerns the magnetic flux tubes constructed from the extrapolation models, henceforth referred to simply as ``flux tubes."
Section~\ref{comparison} presents the comparison of expansion factors measured for the loops and flux tubes. 
Section~\ref{conclusion} is a discussion of the results and their significance.


\section{Observed {\it TRACE} Loops}
\label{loops}

\subsection{Data Description}


Our {\it TRACE} dataset consist of full resolution
(0.5 arcsec $\approx 0.36 Mm$ pixel) images 
obtained in the Fe IX 171 \AA ~passband 
(Handy et al. 1999).  The dates, times, and positions of the images are given in Table~\ref{trace_data}.  These particular images were chosen because clean loops (relatively unobscured by background emission or overlapping loops) could be identified and because the active regions are close to disk center, which is important for the magnetic field modeling.  The data were processed using the standard SolarSoft analysis tools.


We selected 20 loops from 3 active regions for detailed study, as indicated 
in Table~\ref{trace_data}.  
We chose the time difference between consecutive images to be long enough 
that a different configuration of loops is observed.  Therefore, the same 
loop is not chosen in consecutive images of the same active region.
Each loop is fully contained within the active region, which is important for the extrapolation procedure. Upper panels in Figures~\ref{ar_jul} and~\ref{ar_oct} show two of the {\it TRACE} images used. In the lower panels the loops are labeled and corresponding 
model field lines are shown (see Section~\ref{fitting_proc}).

\subsection{Loop Analysis}
\label{obs_loops}


We follow the analysis procedure of Klimchuk et al. (1992).  We start by making a straightened version of the loop.  We visually select a set of 
points that we believe to lie along the loop axis and fit them with a  
polynomial, which then defines one axis of a new coordinate system.  The other axis is orthogonal to it at each axis position.  Intensities are assigned to a regular grid of pixels in the new coordinate system using a weighted average of the four nearest pixels in the original coordinate system.  This results in a rectangular image with the loop running vertically down the middle.  We next determine a background by subjectively outlining the edges of the straightened loop and performing linear interpolations between the intensities at the two edges in each row.  This is the same procedure used on {\it TRACE} loops by Watko and Klimchuk (2000).  Klimchuk et al. (1992) and Klimchuk (2000) used a polynomial surface fitting procedure to determine the background of  {\it Yohkoh} loops, but this can introduce spurious results when the background is highly structured, as is often the case in {\it TRACE} images.  
Finally, we compute the standard deviation (i.e., the second moment) of the intensity profile along each row of the background-subtracted image:
\begin{equation}
\label{sigma}
\sigma = \left[\frac{\sum \left(x_i - \mu\right)^2 I_i}{\sum I_i}\right]^{1/2} ,
\end{equation}
where the summation is taken over the $x_i$ positions in the profile,
and $\mu$ is the mean position:
\begin{equation}
\mu = \frac{\sum x_i I_i}{\sum I_i} .
\end{equation}

It is easy to demonstrate that the standard deviation of
the intensity profile is 1/4 of the loop diameter for the case of a 
circular, uniformly filled cross section and observations with perfect 
resolution. Many authors freely use the term ``diameter" in discussing loop 
thickness, but we are reluctant to do so, because it implies that 
the cross section is necessarily circular.
Instead, we use the term ``width."  To facilitate comparison 
with other published results, we define the width to be 4 times the standard 
deviation. 
The triangles in Figure~\ref{stdev} show the width plotted as a function of position along the loop for four example loops.  The units are Mm in this and all subsequent figures.  Note how the width is fairly constant along each loop.  There is considerable small-scale structure, but little evidence for large-scale trends.

In a recent study of {\it TRACE} loops by Aschwanden and Nightingale (2005), the  intensity profiles are fit to Gaussians with a 
linear background.  For a Gaussian, the full width at half maximum (FWHM) is 
2.35 times as large as the standard deviation, or 0.59 times the 
width as we have defined it.  The mean FWHM of 1.42 Mm measured by 
Aschwanden and Nightingale corresponds to a mean width of 2.41 Mm, which is 
very similar to the values we have measured (e.g., triangles in 
Figure~\ref{stdev}).


Our background subtraction procedure is of course not perfect, and 
residual non-loop emission may be present in the background-subtracted 
images.  Since intensity is multiplied by the square of position in 
Equation~\ref{sigma}, there may be concern that residual emission in the 
tails of the intensity profile could have an especially strong influence on 
$\sigma$.  In the Appendix section we explore this possibility in some 
detail. There we determine the widths of the loops 
of Figure~\ref{stdev} using simultaneous measurements of the standard deviation, 
the FWHM, and the Equivalent Width.
The results obtained using the three different methods are 
very similar.  In particular, fluctuations along the loops, which could 
be an indication of errors in the background subtraction, are of the same 
order. This confirms the suitability of using the standard deviation of 
the intensity profile for width determinations.


The assumption of perfect resolution is of course not realistic.  To quantitatively estimate the effects of finite resolution, we have simulated the observation of a circular cross section loop with a point spread function (PSF) appropriate to the combined telescope/detector system of {\it TRACE}.  Figure~\ref{width_stdev} shows the resulting relationship between the computed standard deviation and the actual loop diameter (width).  The curve rolls over from the straight line of slope 4 and intersects the abscissa at a value of 0.33 Mm, which is the standard deviation of the PSF itself.

The PSF used in Figure~\ref{width_stdev} is a Gaussian with a full width at 
half maximum (FWHM) of 2.25 pixels (approximately 0.82 Mm). 
Golub et al. (1999) applied a 
blind iterative technique to {\it TRACE} images and found that the PSF 
is in fact not azimuthally symmetric. The FWHM of the major and 
minor axes were estimated to be 3 and 2 pixels, respectively.  
The estimate of the major axis was subsequently revised to be 
closer to 2.5 pixels (Golub, 2003, private communication).  
Recent analysis by Nightingale (2003, private communication) 
indicates major and minor axes of only 2.0 and 1.6 pixels.  
We feel that our choice of 2.25 pixels for both axes is 
conservative and, if anything, overestimates the actual smearing 
of the telescope and detector. We note that the conversion curve 
given in Watko and Klimchuk (2000) is not quite correct.  It assumes 
a FWHM of 2.5 pixels for the telescope PSF alone and separately treats 
the averaging over finite pixel area. The authors did not recognize 
at the time that the Golub et al. (1999) result represents a combined 
PSF of the telescope/detector system.

The asterisks in Figure~\ref{stdev} are obtained from the standard deviations using the conversion curve of Figure~\ref{width_stdev}.  We refer to these as resolution-corrected widths.
As expected from the nonlinearity of the conversion curve, 
the difference between the corrected and uncorrected values (asterisks and 
triangles respectively in Figure~\ref{stdev}) is greatest 
for the narrowest loop regions. For measured standard deviations close to the theoretical minimum, the correction can be very large.  It sometimes happens that the measured value is smaller than the theoretical minimum, in which case the width is set 
to zero. See, for example, the upper-right panel in Figure~\ref{stdev}.  We have rejected loops for which 20\% or more of the inferred widths are under the resolution limit. 


We believe that the greatest source on uncertainty in the measurement 
is the subtraction of the background emission from the loop.  
To estimate this uncertainty, our analysis routine automatically 
repeats the width measurement after redefining the edge of the loop to be one pixel wider on each side.  We find that the width obtained in this way 
differs from the original width by about 20\% on average.  Furthermore, 
it is wider than the original width in a large majority of cases.  
We conclude from this that our subjective identification of the loop 
tends to miss the faint outer ``wings" of the profile.

Although we most often underestimate the width, it is nonetheless 
appropriate to consider the possibility of an overestimate when establishing 
the uncertainties.  
An overestimate uncertainty of 20\% seems unreasonably large, so we 
adopt a value of 10\%.  The solid lines in Figure~\ref{stdev} are 
resolution-corrected widths obtained from uncorrected widths that are 20\% higher and 10\% lower 
than the original measured values.  
The lines are 5-point running averages, which makes the plots more 
readable, but does not affect any conclusions since large width variations over short distances are not reliable.  We refer to the lines as the error bars.

\subsection{Loop Expansion Factor}
\label{obs_width}



To quantify systematic variations in loop width, we define an expansion factor $\Gamma$ to be the ratio of the average widths measured in different segments of the loop---at the ends and at the midpoint.  The segments have a length that is 15\% of the total loop length.  We subjectively define the footpoints of the loop to be the locations where the intensity pattern can no longer be confidently identified.  Klimchuk (2000) used a quantitative definition that tends to underestimate the true length.  As we discuss later, the footpoints of the observed loop do not always correspond closely to the locations where the corresponding flux tube intersects the photosphere.
The footpoint segments are labeled ``start" and ``end" for reasons related to the flux tube construction, and the ``middle" segment is exactly halfway between them.  The start footpoint always appears to the left in the figures.  We compute average widths for the start ($W_s$), middle ($W_m$), and end ($W_e$) segments using the resolution-corrected measurements, but ignoring values below the resolution limit because their uncertainties are so large.

We are interested in differences between footpoints in addition to footpoint-midpoint differences.  We therefore define four different expansion factors:

\be
\Gamma^*_{m/s}=\frac{W_m}{W_s},
\ee

\be
\Gamma^*_{m/e}=\frac{W_m}{W_e},
\ee

\be
\Gamma^*_{m/se}=\frac{2~W_m}{(W_s+W_e)},
\ee

\noindent and 
\be
\Gamma^*_{e/s}=\frac{W_e}{W_s},
\ee


\noindent
where the asterisk is used to distinguish these definitions from slightly 
modified definitions introduced in Section 4.
These four expansion factors were computed for each of the 20 loops, and Table~\ref{gamma_table1} shows their means and standard deviations (not to be confused with the standard deviations of the intensity profiles used to determine the widths).  On average, the loops are only about 15\% wider at their middles than at their footpoints, and the start footpoints are statistically about as wide as the end footpoints.  The standard deviations are not small, however.  
These results are entirely consistent with earlier findings on footpoint-to-midpoint expansion factors.  Klimchuk (2000) reported a mean value of 1.30 for a sample of 43 loops observed by {\it Yohkoh}, and Watko \& Klimchuk (2000) reported mean values of 0.99 and 1.13 for, respectively, non-flare and post-flare loops observed by {\it TRACE}.  Klimchuk et al. (1992) found a mean expansion factor of 1.13 for 10 {\it Yohkoh} loops using uncorrected width measurements.


Finally, we test for any dependence of the expansion factor on the 
loop width.  If the observation of nearly uniform width were an artifact of 
poor spatial resolution, then we would expect the expansion factor to be 
positively correlated with the width.  
In Figure~\ref{gamma_width} we plot $\Gamma^*_{m/s}$ versus the average 
width of the middle portion $W_m$. It can be
clearly seen that there is no correlation, which we have verified using 
a nonparametric statistical analysis.


\section{Modeled Magnetic Flux Tubes}
\label{fluxtubes}

\subsection{Linear Force-free Extrapolation}


We extrapolate the observed photospheric field into the corona using the 
method described in Alissandrakis (1981). It employs a
Fast Fourier Transform (FFT) procedure to solve the linear force
free-field equation (Equation \ref{eq:ff}) with $\alpha$ equal to a constant.  
The numerical code was developed by 
D\'emoulin et al. (1997) and has been used in a number of studies (e.g., Green et al. 2002). 
The computational volume is a 3D Cartesian box with the $z=0$ plane corresponding to the photosphere.  Periodic boundary conditions are imposed on the side walls, and the field strength is required to decrease at the limit of large heights. This implies that
each Fourier mode has an exponential decrease in the $z$ direction (with a scale height that depends on both the spatial wavelength of the mode and $\alpha$). 
 The actual calculations are performed in two-dimensional Fourier space with 256$\times$256 horizontal modes. In order to save computer space, the results are kept on a 129$\times$129 nonuniform grid.
 The solution is discretized in 81 steps in the $z$-direction.
 Full-disk longitudinal magnetograms from SOHO/MDI (Scherrer et al. 1995) are used to specify the vertical component of the field  at the lower boundary.  The magnetograms have a spatial resolution of $\approx$ 1.44 Mm/pixel, but we interpolate onto a grid with a somewhat smaller spacing of 1 Mm in the finest part.  To minimize the contribution of unknown transverse 
components to the observed line-of-sight field, we restrict our analysis to 
active regions that are close to disk center.  We choose the horizontal dimensions of the box to be large enough that all of the active region is contained within it.  Any flux imbalance is offset by not taking into account the Fourier mode (0,0) (this correspond to an added uniform weak field). 

\subsection{Image Coalignment}


To study the magnetic properties of individual coronal loops, it is necessary to have an accurate coalignment between the magnetograms and {\it TRACE} images.  
According to the SolarSoft {\it TRACE} Analysis Guide the uncertainty of the {\it TRACE} pointing is 5-10 arcsec, though recent calibrations seem to have improved these numbers (Aschwanden 2005, private communications). Since we consider that this is not adequate for our purposes we use 171 \AA~images from the Extreme-ultraviolet Imaging Telescope (EIT) on SOHO (Delaboudiniere et al. 1995) as an intermediate step. The EIT images and MDI magnetograms are both full disk and can be accurately coaligned by forcing the solar limbs to agree.  Repeated attempts to coalign a single EIT/magnetogram pair suggest an uncertainty of approximately 0.5-1.0 EIT pixel (0.9--1.9 Mm).  

We then coalign the EIT and {\it TRACE} images by matching features that are visible in both.  Visual inspection seems to work better than a purely quantitative cross-correlation approach.  Many 171 \AA~ features are quite stable, such as moss and the footpoints of large-scale magnetic structures (Berger et al. 1999, Martens et al. 2000), but other features evolve significantly over timescales of minutes.  These changes can be identified and ignored in a visual comparison, but they have a significant influence on the cross correlation.  Intensity differences resulting from differences in the {\it TRACE} and EIT bandpasses are also better treated by visual comparison.

Once the {\it TRACE} image is coaligned with the EIT image, it can be straightforwardly coaligned with the magnetogram by accounting for the small offset due to solar rotation during the time lag between the observations.  The time lag is less than one hour, so there is no need to account for the distortions produced by differential rotation.  We estimate an approximate 2 Mm uncertainty in the final {\it TRACE}/magnetogram coalignment.

\subsection{Identifying Magnetic Axes of Loops}
\label{fitting_proc}

The first step in constructing a flux tube model of a loop is to identify the field line at the loop's axis.  The procedure is described in Green et al. (2002).  Essentially, we compute many different linear force-free field models, each characterized by a unique $\alpha$, and compute 
many different field lines to find the one field line that best fits the {\it TRACE} loop as seen projected onto the plane-of-the-sky.  The mean separation between the field line and loop axis is a quantitative measure of the fit. 
More precisely, for each defined point along the loop axis, we first find the closest point in the computed field line. This defines the local separation. Then we obtain the mean separation for all the loop axis points. This procedure permits to find the closest field line without the need to define the end points of the loop.
 For each model we trace many field lines from a square grid of starting positions that is centered on the better defined of the two loop footpoints (the ``start" footpoint). We repeat the procedure with successively finer grids until we find the best fit for that particular model.  We do this for many models covering a range of $\alpha$. The upper panel in Figure~\ref{fitting} shows an example of magnetic axis fitting.  The green box corresponds to the initial sub-grid for the field line tracing, the blue crosses indicate the observed loop axis, and the red line
is the best model field line for the axis. Figure~\ref{alpha} shows the mean separation of the best fit field line plotted as a function of $\alpha$ for one of our loops.  The model with the smallest mean separation (the undimentionned $\alpha$ value is $-0.15$ in this example) is used in the final analysis.  We reject cases where the smallest mean separation is greater than 2 Mm.  Note 
that the linear force-free field provides a much better fit than the potential 
field ($\alpha = 0$).

As we have mentioned, the detectable footpoint of the {\it TRACE} loop is generally offset from the photospheric footpoint of the corresponding flux tube.  We have performed hydrodynamic loop models which suggest that detectable {\it TRACE} emission should be present to within a short distance of 
the chromsphere.   Some of the observed offsets are consistent with the expected 
thickness of the chromosphere (few thousand kilometers for an inclined flux tube), while others are much larger.  In some cases the lower leg is simply
obscured by bright background emission.  In others it is too faint to be readily detected, apparently contradicting the hydrodynamic models.  In still other cases it seems clear that the model field simply does not represent the loop accurately in the vicinity the footpoint.  With some exceptions, the maximum footpoint offset that we allow at either end of the loop is 20 Mm. 
When the offset is greater, we choose a different field line even if the overall fit is not as good (though the mean separation must always be less then 2 Mm).  In a few cases there is reason to believe that the flux tube is considerably longer than the visible loop, such as when the visible end falls in a region of incorrect polarity but points to a region of correct polarity.  We make a subjective decision to keep these cases.  In a small minority of cases both a short and a long field line give acceptable fits, and we keep both.  In no instance, however, is the offset allowed to exceed 60 Mm.  While we believe 
these selection criteria are very reasonable, we present separate results for 
the cases where the offset is less than 10 Mm.  The results are similar.

Figures~\ref{ar_jul} and~\ref{ar_oct} lower panels show (in red) the best-fit field lines overlaid on {\it TRACE} images for a number of loops in two of the studied active regions. Loop 3 in Figure~\ref{ar_jul} and loop 1 in Figure~\ref{ar_oct} are shown also in panels (a) and (d), respectively, of Figures~\ref{stdev},~\ref{width_plots_1}, and~\ref{width_plots_2}. 

\subsection{Constructing Flux Tubes}
\label{construction}


We remind the reader that the term ``flux tube" refers explicitly to 
a magnetic flux tube based on an extrapolation model, and 
the term ``loop" refers explicitly to an EUV intensity feature observed in a 
{\it TRACE} image.  The terms are distinct and not interchangeable. 

Flux tubes are constructed using the best fit field line as the tube axis.  The shape of the tube cross section is unknown and must be assumed at some location along the axis.  We choose to define the shape at the start footpoint.  We consider four possible footpoint shapes, as indicated in Figure~\ref{footpoints}:  a circle and three ellipses contained in the photospheric (x-y) plane.  The major axes of the ellipses are oriented perpendicular to and at $45\deg$ angles to the projection of the tube axis in the x-y plane.  Note that the footpoint shape will be
different from the cross section whenever the tube axis is inclined to the
vertical.  
For each of the footpoint shapes, we trace 24 field lines starting from points distributed systematically around the perimeter.  These field lines define the shape and size of the cross section throughout the remainder of the tube, which can be highly variable. 
The lower panel in Figure~\ref{fitting} shows a flux tube constructed using a circular footpoint centered on the best-fit field line of the upper panel.
We determine an ``observed" width by finding the spread of the field line bundle when viewed in projection onto the plane-of-the-sky.  The width is measured perpendicular to the tube axis.  

Initially, we normalize the flux tubes by setting the width of start footpoint equal to the resolution-corrected width of the loop segment nearest that footpoint, $W_s$ (this segment is 15\% of the total length of the loop, see Section~\ref{obs_width}). The ratios of the major to minor axes of the ellipses are chosen so that all three ellipses have the same area (though different from the circle) and therefore enclose approximately the same magnetic flux. 
Figure~\ref{width_plots_1} shows how the width varies along the tube for the same loops shown in Figure~\ref{stdev}.  The broken thin lines represent the four different cross sections (circle and 3 ellipses).  Also shown as asterisks between solid thin lines are the loop widths and their error bars from Figure~\ref{stdev}.  Note that the vertical scale in Figure~\ref{width_plots_1} is greatly expanded compared to Figure~\ref{stdev}.  In general the flux tubes are much wider than the loops everywhere except the start footpoint, where the normalization forces them to be similar.  The flux tubes also tend to be longer than the loops, especially on the ``end" side, for the reasons we have discussed.


The thick solid lines in Figure~\ref{width_plots_1} indicate the square root of the cross-sectional area, $A^{1/2}$, obtained from the on-axis field strength using conservation of magnetic flux: $\Phi = B \; A$ (i.e., we plot $B^{-1/2}$).  This is strictly correct only for very thin flux tubes in which $B$ is constant over the cross section.  We have normalized $A^{1/2}$ in the same way as the flux tubes.  Roughly speaking, $A^{1/2}$ represents the average width that would be measured if the loop were observed over a complete $360\deg$ range of viewing angles.  This is in contrast to the single viewing angle represented by the flux tube width curves in the figure.  Large differences between the $A^{1/2}$ and flux tube width curves (e.g., panel c) indicate highly non-circular cross sections, because the widths of such cross sections are strongly dependent on the viewing angle.  A good example of this is shown in Figure~\ref{example}.  The top panel shows the flux tube as viewed from above.  This is the
same flux tube that gives a reasonably good 
fit to the {\it TRACE} loop in panel d of Figure~\ref{width_plots_1} (dot-dash curve).  The bottom panel shows the same flux tube as viewed from the side.  The difference is dramatic.


To better compare the flux tubes and loops, we consider a second normalization that forces the average width of the tube and loop to be similar over the entire region of overlap, i.e., over the entire loop except in the rare instances where the loop extends beyond the flux tube (e.g., panel c).  To obtain the new 
normalization, we first determine the factor by which the original 
tube must be narrowed in order for its average width to equal the average width 
of the loop.  We then shrink the cross section of the start footpoint by this 
factor and trace a new bundle of field lines. Since this is a new flux 
tube (and not simply a reduced-width version of the original), its width is 
slightly different from the width of the original tube reduced by the factor. 
The width curves of the tubes 
obtained with the second normalization are shown in Figure~\ref{width_plots_2} using the same format as Figure~\ref{width_plots_1}.  
The agreement with the loop widths is improved, but still quite poor in many cases.  In some instances there is at least one flux tube width curve that falls mostly within the loop width error bars (e.g., panels b and d).  In other instances there is gross deviation for all of the curves (e.g., panels a and c).  Even in the cases with reasonable agreement, there is a tendency for the flux tube widths, but not the loop widths, to vary systematically along the loop.  In panel b, for example, the flux tube widths are systematically higher on the right side than on the left side, and in panel d they are systematically higher in the middle.  The loop widths do not exhibit the same trends.  These differences will be more apparent when we compare expansion factors for the tubes and loops in the next section.


Before proceeding, we examine the possibility that the model flux tubes we have identified with loops do not accurately represent the actual flux tubes on the Sun.  In most cases the best-fit model field line used for the tube axis coincides quite well with the observed loop, but the match is never perfect.  In a few cases the match seems qualitatively rather poor even though the fit criteria of Section 3.3 are satisfied.  Differences arise for two primary reasons:  the uncertainty in the MDI-{\it TRACE} coalignment and the assumption of a linear force-free field.  The force-free approximation is probably quite good, so $\alpha$ will be constant on each field line, but it is unreasonable to expect $\alpha$ to be the same on all field lines.  In more realistic {\it non}linear force-free fields, $\alpha$ varies across the active region.  Even if the $\alpha$ of our model is the actual $\alpha$ of the loop axis field line, the shape of the field line will depend on the distribution of $\alpha$ elsewhere in the active region.  We would not expect the flux tube and loop to coincide precisely.  To evaluate the importance of these effects, we investigated one representative case in considerable detail.  We repeated our analysis multiple times by introducing MDI-{\it TRACE} offsets of 2 Mm in all four compass directions, and by constructing flux tubes using $\alpha$ values that differ by 20\% from the best-fit $\alpha$. The resulting changes are of the order of 10\% for $A^{1/2}$ and no more than 20\% for the width.  We therefore believe, but cannot prove definitively, that the sizable discrepancies we have found between flux tubes and loops are real. 


\section{Comparing Flux Tube and Loop Expansion Factors}
\label{comparison}

To quantify the systematic differences between flux tubes and loops, we compute expansion factors using precisely the same segments from both structures.  We modify the segment definitions of Section~\ref{obs_width} to take into account the flux tubes as much as possible, which we consider to be the more fundamental structures.  As we have discussed, parts of many flux tubes are not easily identifiable as loops because the loop emission is weak or the background emission is strong.  Panel (a) in Figures~\ref{width_plots_1} and~\ref{width_plots_2} is an example of the problems that arise when using the earlier definitions.  The end segment of the loop is closer to the midpoint of the flux tube than it is to the right footpoint.  It is more properly classified as a middle segment than an end segment.  As before, all of the segments in our new definitions have a length that is 15$\%$ of the loop length.  Also as before, the start segment is the first 15$\%$ of the loop---the part nearest the footpoint where we define the cross section and begin the field line traces.  There is usually good correspondence between the loop and flux tube on this end.  If the loop extends beyond the midpoint of the flux tube, the middle segment is now defined to be the 15$\%$ segment centered on the {\it tube} midpoint, rather than the loop midpoint, and the end segment is the last 15$\%$ of the loop.  If the loop does not reach the midpoint of the flux tube, the last 15$\%$ of the loop is assigned to the middle segment, and there is no end segment.

We compute $\Gamma_{m/s}$, $\Gamma_{m/e}$, and $\Gamma_{e/s}$ expansion factors from the average widths in these segments in the same way as before.  The asterisk has been removed to indicate that the segment definitions are different from those in Section~\ref{obs_width}.
We use flux tube widths based on the second normalization.  Because the segments are sometimes far removed
from the footpoints or midpoint of the flux tube, we reject cases that do not satisfy certain constraints.  The constraints are based on the separation between the centers of the segments as measured along the flux tube axis.  $\Gamma_{m/s}$ and $\Gamma_{m/e}$ are included only if the separation between the middle segment and the start or end segment, respectively, is greater than 25$\%$ of the {\it flux tube} length.  $\Gamma_{e/s}$ is included only if the separation between the start and middle segments and the separation between the middle and end segments are both be greater than 15$\%$ of the flux tube length.  

Each loop is associated with four flux tubes, one for each cross section 
(circle and three ellipses).  A small subset of loops have eight flux tubes because two different best-fit field lines are acceptable.  For each loop, we average together the flux tube expansion factors of each type to obtain composite $\Gamma^{tube}_{m/s}$, $\Gamma^{tube}_{m/e}$, and $\Gamma^{tube}_{e/s}$.  We also compute $\Gamma^{A^{1/2}}_{m/s}$, $\Gamma^{A^{1/2}}_{m/e}$, and $\Gamma^{A^{1/2}}_{e/s}$ using the square-root of the cross sectional area (we actually use $B^{-1/2}$).  Finally, we have $\Gamma^{loop}_{m/s}$, $\Gamma^{loop}_{m/e}$, and $\Gamma^{loop}_{e/s}$ from the resolution-corrected {\it TRACE} measurements.  We emphasize that the corresponding flux tube and 
loop expansion factors are computed using the same intervals of $s$ (see Figure~\ref{width_plots_2}), so that the comparison is meaningful.

It is appropriate to average the results for the different cross sections 
because this is a statistical study of many loops.  If we were investigating a 
single loop and found that a particular cross section is able to reproduce the loop much better than the others, then we could be plausibly conclude that 
it is the actual cross section.  It would not be unreasonable to suggest that 
coronal heating selects that particular bundle of field lines to be illuminated.  
However, it is beyond common sense to suggest that coronal heating always picks the right bundle of field lines for each of a large collection of loops.  That would imply that there is something special about the position of the 
{\it TRACE} spacecraft with respect to the Sun. (This is even true of the 
circular footpoint, since the circle becomes distorted along the 
loop and the symmetry is broken.)  We therefore choose to average the expansion factors over the four assumed cross sections.  We note 
from Figures~\ref{width_plots_1} and~\ref{width_plots_2} that the expansion 
factors are not qualitatively different for the different footpoint cross sections.  

Figure~\ref{gammas} shows scatter plots of $\Gamma^{tube}_{m/s}$ versus $\Gamma^{loop}_{m/s}$, $\Gamma^{tube}_{m/e}$ versus $\Gamma^{loop}_{m/e}$, and $\Gamma^{tube}_{e/s}$ versus $\Gamma^{loop}_{e/s}$. Asterisks correspond
to cases in which the loop footpoint and the flux tube footpoint are separated by
less than 10 Mm (pluses correspond to the rest of the cases, i.e., footpoint separation larger than 10 Mm). It can be seen that restricting the distance between loop and flux tube footpoints (see Section~\ref{fitting_proc}) does not affect the qualitative properties of the scatter distribution.
Table~\ref{gamma_table2} gives the mean and standard deviation statistics on the different expansion factors. The figure and table reveal a number of interesting properties. 
First consider the behavior of the loops.
If we imagine vertical lines at $\Gamma^{loop}=1.0$ in Figure~\ref{gammas}, we see that the points are divided roughly in half in all three panels.  Just as many loops exhibit footpoint-to-midpoint constriction as exhibit footpoint-to-midpoint expansion.  Neither the constriction nor the expansion is very large (typically less than 50\%).  Half of the loops are wider at their left footpoint than at their right footpoint, and {\it vice versa}.   Not surprisingly, the mean values of the three expansion factors are only slightly larger than 1.0 in Table~\ref{gamma_table2}.  The results are similar to those of Table~\ref{gamma_table1}, based on the original loop segment definitions, and are fully consistent with previously published results.

Now consider the behavior of the flux tubes. 
Imagine horizontal lines at $\Gamma^{tube}=1.0$ in Figure~\ref{gammas}.  All but two of the points in the upper panel lie above the line, indicating an overwhelming tendency for flux tubes to expand from the start footpoint to the midpoint.  $<\Gamma^{tube}_{m/s}> = 1.79$, so the tubes are nearly twice as wide at their midpoints as at their start footpoints, on average.  This is in sharp contrast to the behavior of loops.  Interestingly, the points in middle panel are nearly evenly divided above and below the imaginary line, so there is no significant preference for expansion or constriction with respect to the end footpoint.
Twice as many points in lower panel lie above the imaginary line as below, indicating a strong tendency for the end footpoint to be wider than the start footpoint.  The difference can be very large, with $<\Gamma^{tube}_{e/s}> = 2.62$.  The extreme asymmetry of flux tubes contrasts sharply with the symmetric nature of loops.

The flux tube asymmetry is much less pronounced in cross-sectional area.  $<\Gamma^{A^{1/2}}_{e/s}> = 1.24$, not much greater than unity, and the standard deviation is only 0.50.  This is a clear indication that the shape of the cross section at the end footpoint is much more irregular and much less compact than the circle and ellipses assumed at the start footpoint.  Field strengths are comparable at the two ends, since they scale directly with the cross-sectional area.

Statistically, loops and flux tubes have very different expansion properties.  These same differences show up clearly in a case-by-case analysis.  We compute the ratios of the flux tube and loop expansion factors for each case separately:  $R_{m/s} = \Gamma^{tube}_{m/s}/\Gamma^{loop}_{m/s}$, $ R_{m/e} = \Gamma^{tube}_{m/e}/\Gamma^{loop}_{m/e}$, and $ R_{e/s} = \Gamma^{tube}_{e/s}/\Gamma^{loop}_{e/s}$.  The last column in Table~\ref{gamma_table2} gives the means and standard deviations of these ratios.  They confirm what we found above:  (1) flux tubes expand twice as much as loops from start footpoint to midpoint; (2) flux tubes and loops expand or constrict to an equal degree from end footpoint to midpoint; and (3) flux tubes are far more asymmetric than loops.  It may seem odd that the $<R_{e/s}>$ is noticably larger than the $<R_{m/s}>$ at the same time that the $<R_{m/e}>$ is essentially unity.  This is not a real inconsistency and arises because the samples are not the same for the different ratios due to our selection criteria.  There are fewer $\Gamma_{e/s}$ cases than $\Gamma_{m/s}$ cases, and still fewer $\Gamma_{m/e}$ cases.  The solid lines in Figure~\ref{gammas} have slopes equal to the mean values of the ratios, while the dotted lines have slopes equal to the mean values plus and minus the standard deviations.  The dashed lines have a slope of 1.  It is important to note that a large majority of the points in upper panel lie well above the dashed line, and only two points in lower panel lie significantly below the dashed line.  Therefore, the trends we have identified are actual trends obeyed by a majority of loops and not simply artifacts of a few unusual (``outlier") cases.

Table~\ref{gamma_table3} gives the statistical results, in the same 
format as Table~\ref{gamma_table2}, for the 
cases in which the loop endpoint and flux tube footpoint are separated by 
less than 10 Mm (asterisks in Figure~\ref{gammas}).  
Possible reasons for the separation are discussed in 
Section 3.3.  We see that the results are very similar, indicating that our 
full sample is not contaminated by bad cases.


\section{Discussion}
\label{conclusion}

Our study demonstrates that linear force-free extrapolation models based on MDI magnetograms are inconsistent with {\it TRACE} loop observations.  The loops have a nearly uniform width, whereas the corresponding flux tubes generally do not.  Most of the flux tubes expand appreciably with height from either one or both footpoints.  Thus, it is {\it not} the case that loops correspond to a subset of uniform-width flux tubes that may exist in magnetic field configurations of the type we have modeled.  This is an important result that had not been ruled out by previous studies.

Another very important result is that the model flux tubes are much less symmetric than observed loops.  Most flux tubes appear considerably wider at one footpoint than the other when viewed from the {\it TRACE} line-of-sight.  The 
cross-sectional areas of the two footpoints are very similar, on the other hand. This indicates that the asymmetry is one of shape rather than field 
strength.  The simple and compact cross sections (circle or ellipse) that we have assumed at the start footpoints map to much more irregular and distended cross sections at the end footpoints.  It is especially interesting that this result is independent of the side on which the field line traces are begun.  When we begin with circles and ellipses at the location of the original end footpoint, we find that they map to irregular and distended cross sections at the location of the original start footpoint.  The flux tubes are asymmetric in both cases, but in an opposite sense.

We can understand this result in terms of two effects.  First, there is a 
general tendency for field strength to decrease with height 
and for cross-sectional area to increase with height to conserve flux.  
Second, there is a tendency for 
field lines to deviate from each other as they are traced and for the 
shape of the cross section to become progressively distorted.
Such distortion is typical in the mapping of field lines in traditional magnetic 
field models 
and can be severe at quasi-separatrix layers (QSLs; see e.g. Titov et al. 
2002).  An important point is that the distortion is independent of the 
direction in which the field lines are traced.  Therefore, when we start 
from a compact footpoint and trace field lines up to the tube apex, both 
the tendency for field strength to decrease and the tendency for the cross 
section to distort cause the width of the tube to increase.  As we continue 
the trace down to the end footpoint, the two effects are competing.  
Continued distortion causes the width to increase, but increasing field 
strength causes the width to decrease.  Which effect wins out varies from 
tube to tube, and $\Gamma^{tube}_{m/e}$ values greater than 1 and less than 
1 are equally common (see Figure~\ref{gammas}).

We note that our decision to start the field line traces at a footpoint is 
merely a technical convenience.  There is no physical motivation.  Which field 
lines are illuminated to form an observed loop is determined entirely by 
coronal heating.  If coronal heating were known to occur at the top of a loop, 
it would be appropriate to trace field lines starting at the top, using the 
cross section defined by the heating.  It is interesting to conjecture that 
flux tubes defined by a circular cross section at the top would tend to be 
more symmetric than flux tubes defined by a circular cross section at one of 
the footpoints.  In the first case, the cross sectional distortion will tend to 
increase while tracing down both legs.  In the latter case, it will increase going up one leg and increase further going down the other.

These fundamental differences between the loops and flux tubes imply that either the plasma structures revealed by {\it TRACE} do not follow magnetic field lines or the magnetic field models we have used do not accurately represent the detailed properties of the solar magnetic field.  The first possibility seems highly implausible.  Thermal conduction and plasma motions are extremely efficient at transporting energy and matter along the magnetic field, but cross-field transport is greatly inhibited (e.g., Litwin \& Rosner 1993).  We conclude that the models are inadequate.  Either the linear force-free approximation is poor or the magnetogram boundary conditions are lacking, or both.

Measurements of the plasma pressure suggest that the field is close to force free throughout most of the corona of active regions.  Even if it were not, plasma pressure effects could not explain the discrepancies we have found.  We believe that the fundamental source of the discrepancies is the assumption of a linear force-free field.  The linear assumption is reasonable for modeling the large-scale structure of the field that is determined by large-scale currents (e.g., the shapes of loop axes, as discussed at the end of Section 3.4), but it cannot treat the small-scale currents that are critical for loop cross sections. Even nonlinear force-free field models based on today's modest resolution magnetograms are inadequate for this purpose.  
The possibility considered by Klimchuk et al. (2000) that loops are locally twisted magnetic flux tubes would imply currents with a transverse scale smaller than a loop diameter.  The nonlinear force-free models that they constructed suggest that twist cannot produce the degree of thickness uniformity observed in real loops; however, the idea cannot be ruled out entirely because the models have a maximum twist of $2\pi$.  Greater twist is likely to produce greater uniformity, but it is not known how much greater the twist can be before realistic curved loops become kink unstable (see Gerrard, Hood, \& Brown 2004).

The twisted loops considered by Klimchuk et al. have a well organized and relatively simple internal structure.  Real loops are likely to be much more complicated.  High resolution observations of the photospheric magnetic field reveal that is clumped into small and intense kiloGauss flux tubes (see Solanki 1993).  The magnetic flux contained in each elemental tube is so small that a single {\it TRACE} loop must contain tens to hundreds of them (e.g., Priest, Heyvaerts, \& Title 2002).  The footpoints of these tubes are randomly displaced about the solar surface by the changing convective flow pattern (e.g., Schrijver et al. 1997).  We therefore expect the field within a loop to be highly tangled, with the elemental strands wrapping around each other in complicated ways.

The basic picture of tangled field was first proposed by Parker (1988).  He suggested that the energy contained in the magnetic stresses associated with the tangling would be liberated in the form of nanoflares.  From energy balance considerations we can conclude that the nanoflares must occur when the angle between misaligned elemental flux tubes is approximately $50\deg$.  (Parker stated this result in terms of a $\sim 25\deg$ tilt from vertical at the base of the corona).  Recently, Dahlburg, Klimchuk, \& Antiochos (2003, 2005) demonstrated that a mechanism called the secondary instability ``switches on" when the misalignment angle reaches this critical value.  They showed that energy is released impulsively and is adequate to heat the corona.  This agrees nicely with studies showing that the density and temperature properties of coronal loops, especially {\it TRACE} loops, are best explained if loops are modeled as collections of unresolved impulsively-heated strands (Cargill 1994; Warren, Winebarger, \& Mariska 2003; Klimchuk 2004).

The concept of internal tangling within loops may also explain our result that loops are much more symmetric than linear force-free field models would predict.  To see how this might be, imagine a field that is initially very simple, so that flux tubes have compact (e.g., circular) cross sections at both ends.  Systematic motions can rearrange the photospheric footpoints, even while maintaining the same overall flux distribution, so that the flux tubes become highly asymmetric.  Suppose, however, that small-scale random displacements are superposed on the systematic flow pattern.  The footpoints of any two elemental flux strands that are initially close together will then separate according to a random walk.  As they do, they will become tangled with other strands.  The tangling can only proceed so far before the secondary instability causes adjacent strands to reconnect, thereby decreasing the level of stress.  For a random walk step size of 1 Mm corresponding to a granulation cell diameter, a loop length of 100 Mm, and a critical angle of $50\deg$, the footpoint separation would not be expected to exceed 5 Mm.  This would seem to preclude 
the possibility of flux tubes that are highly asymmetric.  Two elemental strands that are close together at one end cannot be greatly separated at the other end.  We plan to investigate this interesting conclusion more thoroughly in future work.  
It is interesting to speculate that the small-scale structure in the measured loop width (e.g., Figure~\ref{stdev}) may be due to irregular trajectories of the strands in the tube bundle.  We are reluctant to make this claim just yet, since 
we cannot rule out the possibility that the ``bumps" in the width curves are 
caused by variable errors in the background subtraction.  The width variations 
along individual loops have a standard deviation that 
is 26\% of the mean width, on average.  This includes systematic variations as well as small-scale fluctuations.  Aschwanden and Nightingale (2005, Figure 9) 
report a standard deviation of 23\%, but their value is artificially small 
because they do not correct for the finite spatial resolution of the 
observations, which is a non-linear correction.

We end on something of a cautionary note.  It was long ago suggested that loops appear to be uniformly wide simply because they are unresolved.  Indeed, if a loop is everywhere narrower than a resolution element (combined point spread function and detector pixel), then its apparent width will be nearly constant even if it has a very large expansion factor.  We have devoted enormous time and energy to addressing this possibility.  Our careful analysis based on the best available information on the point spread functions of {\it TRACE} and {\it Yohkoh}/SXT indicates that both instruments are able to resolve the envelope of 
emission that we identify as a loop.  Of course there could be unresolved internal structure.  There are abundant examples of features in both data sets 
that are as small as the point spread functions measured before launch, indicating that the instruments are performing as expected.  Furthermore, as discussed in Section 2.3, we find no evidence for a correlation between 
expansion factor and width as would be expected if the loops were poorly 
resolved.

It is nonetheless somewhat unnerving that the both {\it TRACE} and {\it Yohkoh} 
loops tend to be a few resolution elements wide, despite the significant 
difference in resolution.  This could be because the $\sim 1$ MK loops detected by {\it TRACE} are physically quite different from the $\sim 2-8$ MK loops 
detected by {\it Yohkoh}.  
There may also be a selection bias at work.  When choosing loops for 
detailed analysis, one is drawn to examples that stand out from the background 
and appear monolithic (i.e., that are not obviously multiple loops).  This 
naturally favors loops that are a few resolution elements wide.  Thinner loops have a reduced brightness contrast relative to the background, especially 
if they are more narrow than a pixel, because then the pixel brightness is an 
average of the intrinsic brightness of the loop and the background.  Many loops 
may actually be small collections of thinner loops, say, 2-5 thinner loops, each thin loop being itself comprised of many kiloGauss flux strands (see also 
Aschwanden, 2005).  The 
collection will appear monolithic if the loops are closely spaced, but not 
if the gaps are comparable to a resolution element.  We therefore 
suggest a picture in which the {\it Yohkoh} loops of our earlier 
studies are actually small collections of unresolved {\it TRACE}-size loops.  
Winebarger \& Warren (2005) have shown that at least some hot {\it Yohkoh} 
loops contain several thinner and cooler {\it TRACE} loops within their 
envelope.  This may not be common (e.g., Schmieder et al., 2004), but it 
suggests that hot plasma may also be structured in thin 
loops.  Whether this is actually the case must 
await high-resolution, high-temperature observations such as may be
available in the mid term from NRL's {\it VERIS} rocket experiment and in the long term from the {\it Reconnection and Microscale (RAM)} and {\it Solar Orbiter} missions.  The AIA instrument on the {\it Solar Dynamics Observatory} 
may also provide useful information on this question, though it remains to 
be seen whether the temperature discrimination will be adequate.


\acknowledgements

We wish to thank Harry Warren and Amy 
Winebarger for useful discussions on {\it TRACE} data processing and 
the instrument pointing. We acknowledge the {\it TRACE} and 
SOHO teams.  We also thank the referee, Markus Aschwanden, 
for his useful suggestions and for encouraging 
the comparison of width measurement techniques that is discussed in the 
Appendix. This work was supported by NASA and the Office of 
Naval Research. 

\clearpage


\appendix
\section*{Appendix: Comparison between different methods of 
loop width determination}
\label{test}

As discussed in Section~\ref{obs_loops} the computation of the
standard deviation of the intensity profile ($\sigma$) implies a 
weighting of the intensity with the square of the position
along the profile (see Equation~\ref{sigma}). It can be argued that this
weighting may artificially amplify the effect of a residual (unsubtracted) 
background at the ``tails'' of the profile.
To explore this possibility, here we compare $\sigma$ with two other 
measures of loop thickness: the Full Width at Half Maximum (FWHM) and
the Equivalent Width ($W_{eq}$), defined as

\begin{displaymath}
W_{eq} = \frac{\sum I_i}{I_{max}},
\nonumber
\end{displaymath}

\noindent
where $I_{max}$ is the maximum intensity along the profile.
These two alternative methods would seem to be less susceptible to 
the effects of residual background.  As we have indicated in Section 2, 
the width (diameter) of a uniformly filled circular cross-section is 
is 4 times $\sigma$.
It can be easily demonstrated that the width is also 1.41 times the FWHM
and 1.27 times $W_{eq}$.

We perform simultaneous measurements of $\sigma$, FWHM, and $W_{eq}$ 
for the four loops shown in Figures~\ref{stdev}, \ref{width_plots_1}, 
and~\ref{width_plots_2}.  We then convert to width using the conversion 
factors above. Since it is not relevant for the present comparison, we do 
not correct the results for finite resolution (see Section~\ref{obs_loops}).
As examples, in Figure~\ref{stdev_test} we plot the 
inferred widths for the loops shown in panels c and d of the figures. 
Clearly, the three methods give similar results, and the fluctuations along 
the loops are of the same order. The computed average fluctuations of 
$\sigma$ for the four analyzed cases are of the order of 15\%. In
comparison, the average fluctuations of the resolution corrected values 
(asterisks in Figures~\ref{stdev}, \ref{width_plots_1}, 
and~\ref{width_plots_2}) are of the order of 26\%. The reason for this
is the non-linearity of the resolution correction curve 
(see Figure~\ref{width_stdev}) for loops near the resolution limit.

It is worth noting that since the results shown in Figure~\ref{stdev_test}
are not corrected for finite resolution they correspond
to the results plotted with triangles in Figure~\ref{stdev} (panels c and d). 
It can be noticed that the 
$\sigma$ measurements in Figure~\ref{stdev_test}
(continuous lines) do not coincide exactly with those in 
Figure~\ref{stdev} (triangles).  Very small differences occur because 
the two figures were made using different measurements
of the same loops, and since our procedure involves two subjective
steps (identifying the loop axis and tracing the loop boundary).  The 
similarity of the measurements indicates that the subjectivity is  
not critical.

In Figure~\ref{linear_fit} we plot $\sigma$ versus FWHM and 
$W_{eq}$ (upper and lower panels, respectively) for the four loops. 
It can be seen that there is a strong statistical correlation
between $\sigma$ and the other two width measures.
We perform least square fits of the scatter data, and we find a 
slope of 0.31 for $\sigma$ as a function of FWHM and 0.35 for $\sigma$ 
as a function of $W_{eq}$. In comparison, the proportionality factors 
based on the approximation of a uniformly filled circular-cross section
are: $\sigma = 0.35$ FWHM and $\sigma = 0.32~W_{eq}$. 
The $\chi^{2}$ test gives a correlation probability of 1.
 
The above results suggest that a circular cross-section loop with uniform
density (except perhaps on a sub-resolution scale) is a reasonable 
approximation, and confirm the suitability of 
the standard deviation of the intensity profile for loop width 
determinations. 

\clearpage


\clearpage

\begin{table}    
\centering
\caption{Description of the {\it TRACE} data used. Date and universal time
(UT) of the images and longitude and latitude of the active region (AR)
are given. The last column indicates the number of observed loops 
selected on each image.}
\label{trace_data}
\vspace{0.5cm}
$\begin{array} {cccc}
 $Date$  & $Time (UT)$ & $AR position$ & $Number of Loops$ \\
\hline
 $Jul-26-2002$ & 07:24 & $W03 N03$ & 5 \\
 $Jul-26-2002$ & 08:33 & $W03 N03$ & 6 \\
 $Jul-29-2002$ & 20:30 & $W23 S18$ & 3 \\
 $Jul-30-2002$ & 01:45 & $W23 S18$ & 2 \\
 $Oct-01-2002$ & 00:30 & $W02 N10$ & 1 \\
 $Oct-01-2002$ & 01:37 & $W03 N10$ & 3 \\
\hline
\end{array}$
\end{table}

\begin{table}       
\centering
\caption{Statistics for the loop expansion factors 
defined in Section~\ref{obs_width}.}
\label{gamma_table1}
\vspace{0.5cm}
$\begin{array} {cc}
         &  $~Mean~~~St. dev.$ \\
\hline
 \Gamma^*_{m/s}  & 1.13~~~~~~0.51 \\
 \Gamma^*_{m/e}  & 1.17~~~~~~0.41 \\
 \Gamma^*_{m/es} & 1.08~~~~~~0.32 \\
 \Gamma^*_{e/s}  & 1.06~~~~~~0.58 \\
\hline
\end{array}$
\end{table}

\begin{table}      
\centering
\caption{Statistics on the redefined expansion factors, $\Gamma$, 
for observed loops, model flux tubes, and square root of the 
cross-sectional area.}
\label{gamma_table2}
\vspace{0.5cm}
$\begin{array} {ccccc}
           &  $Loops$ & $Flux Tubes$ & A^{1/2} & 
R=\Gamma^{tube}/\Gamma^{loop} \\  
\hline
         &  $~Mean~~~St. dev.$ &  $~Mean~~~St. dev.$ &  $~Mean~~~St. dev.$ & $~Mean~~~St. dev.$ \\
\Gamma_{m/s}  &  1.09~~~~~~0.43  &  1.79~~~~~~0.69 & 1.70~~~~~~0.50 &   2.02~~~~~~1.29  \\
\Gamma_{m/e}  &  1.19~~~~~~0.52  &  1.23~~~~~~0.90 & 1.54~~~~~~0.71 &   1.01~~~~~~0.54  \\ 
\Gamma_{e/s}  &  1.35~~~~~~0.82  &  2.62~~~~~~2.53 & 1.24~~~~~~0.50 &   2.70~~~~~~3.53  \\
\hline
\end{array}$
\end{table} 

\begin{table}      
\centering
\caption{Idem Table~\ref{gamma_table2} for cases in which loop and 
flux tube footpoints are separated by less than 10 Mm (see 
Section~\ref{comparison}).
}
\label{gamma_table3}
\vspace{0.5cm}
$\begin{array} {ccccc}
           &  $Loops$ & $Flux Tubes$ & A^{1/2} & 
R=\Gamma^{tube}/\Gamma^{loop} \\  
\hline
         &  $~Mean~~~St. dev.$ &  $~Mean~~~St. dev.$ &  $~Mean~~~St. dev.$ & $~Mean~~~St. dev.$ \\
\Gamma_{m/s}  & 1.22~~~~~~0.47 & 1.65~~~~~~0.67 & 1.44~~~~~~0.25 & 1.62~~~~~~0.95  \\
\Gamma_{m/e}  & 1.19~~~~~~0.52 & 1.24~~~~~~0.90 & 1.54~~~~~~0.71 & 1.01~~~~~~0.54  \\
\Gamma_{e/s}  & 1.42~~~~~~0.92 & 2.46~~~~~~2.89 & 1.13~~~~~~0.49 & 2.60~~~~~~4.04  \\
\hline
\end{array}$
\end{table}

\begin{figure*}   
   \centering
  \hspace{0cm}
\includegraphics[bb= 120 90 460 770,width=10.cm]{f1.eps}
      \caption{Upper panel: \textit{TRACE} image of one of the studied active regions. The lower panel shows the labeled loops and the corresponding best-model field lines (see Section~\ref{fitting_proc}). Units are Mm. Isocontours correspond to 50, 200 and 500 Gauss levels of the longitudinal component of the photospheric magnetic field (outward in green, inward in pink).}
         \label{ar_jul}
\end{figure*}

\begin{figure*}   
   \centering
  \hspace{0cm}
\includegraphics[bb= 120 90 460 770,width=10.cm]{f2.eps}
      \caption{Idem Figure~\ref{ar_jul} for the active region observed on October
1 2002.}
         \label{ar_oct}
\end{figure*}

\begin{figure*}   
   \centering
  \hspace{0cm}
\includegraphics[bb= 25 230 550 635,width=16.5cm]{f3.eps}
      \caption{Plots of width (diameter) versus position along the loop, $s$, for four example {\it TRACE} loops.  Triangles 
correspond to the uncorrected widths, defined as the measured standard deviations of the intensity profiles
multiplied by 4. Asterisks correspond to the resolution-corrected
width values obtained using the curve in Figure~\ref{width_stdev}.
Solid lines are 5-point running averages of resolution-corrected width values 
that account for +20\% and -10\% uncertainties in the uncorrected widths. These lines define estimated error bars for the resolution-corrected widths, as explained in Section 2.2. Dates and times of the observations appear on the tops
of the panels. }
         \label{stdev}
\end{figure*}

\begin{figure*}   
   \centering
  \hspace{0cm}
\includegraphics[bb= 55 360 559 720 ,width=16.5cm]{f4.eps}
      \caption{Relationship between actual width (diameter) and standard deviation of the cross-axis intensity profile for a simulated observation of a loop with a circular, uniformly-filled cross section (see Section~\ref{obs_loops}).  An instrumental PSF with a FWHM of 2.25 pixels 
(0.82 Mm) has been assumed. A dotted line of slope 4 is shown for comparison.}
\label{width_stdev}
\end{figure*}

\begin{figure*}   
   \centering
  \hspace{0cm}
\includegraphics[bb= 85 370 545 695, width=16.5cm]{f5.eps}
      \caption{Expansion factor $\Gamma^*_{m/s}$ versus the average width of 
the middle portion of observed loops ($W_m$). The plot shows no correlation 
between expansion and width (diameter) in the studied loop set.}
\label{gamma_width}
\end{figure*}

\begin{figure*}   
   \centering
  \hspace{0cm}
\includegraphics[bb= 120 90 460 770 ,width=9.cm]{f6.eps}
      \caption{Example of the fitting procedure (upper panel).
Blue crosses are selected points along the loop that are used to compute the
best fit magnetic field line (in red). The rectangle marks
the area from which a number of field lines are traced until the best
fit is reached through a refinement process (see Section 3.3).
Flux tubes (lower panel) are constructed by tracing field lines around the
best fit line (in the present example the footpoint is circular with 
a diameter of 0.8 Mm). }
         \label{fitting}
\end{figure*}

\begin{figure*}   
   \centering
  \hspace{0cm}
\includegraphics[bb= 80 360 560 710 ,width=14.cm]{f7.eps}
      \caption{Example of mean separation between loop axis and best fit field line versus $\alpha$.  $\alpha$ is given in units of $2\pi$/(100Mm). The solid line is a cubic interpolation of the points (asterisks).}
         \label{alpha}
\end{figure*}

\begin{figure*}   
   \centering
  \hspace{0cm}
\includegraphics[bb= 60 180 570 680 ,width=16.cm]{f8.eps}
      \caption{Examples of shapes and orientations of flux tube footpoints.
The x-y plane corresponds to the photosphere.
The crosses mark the start points of field lines to be integrated to
construct the flux tubes. The arrows indicate the projection of the flux tube axis
(the best fit field line) on the x-y plane. In the examples a width of
2 Mm is assumed for the flux tube footpoints.}
         \label{footpoints}
\end{figure*}

\begin{figure*}   
   \centering
  \hspace{0cm}
\includegraphics[bb= 25 215 570 650 ,width=16.5cm]{f9.eps}
      \caption{Examples of width (diameter), indicated by discontinuous lines, and square-root
of the cross-sectional area, indicated by thick solid lines, for flux tubes obtained with 
the first normalization in which the tube width at the start footpoint is set equal to the width of the nearest loop segment. The discontinuous lines 
correspond to the different footpoint shapes as follows: circle (dot), 
perpendicular ellipse (dash), ellipse at 45 degrees (dash-dot), 
ellipse at 135 degrees (dash-triple dot) 
(see Figure~\ref{footpoints}). 
Asterisks and thin solid lines correspond to the same 
data shown in Figure~\ref{stdev}. The coordinate $s$ represents distance 
along the projection of the flux-tube axis onto the plane-of-the-sky
(the origin of $s$ is shifted compared to Figure 3 where $s$ is running along
the corresponding loop, see Section 3.4 for details).
}
         \label{width_plots_1}
\end{figure*}

\begin{figure*}   
   \centering
  \hspace{0cm}
\includegraphics[bb= 25 215 570 650 ,width=16.5cm]{f10.eps}
      \caption{Similar to Figure~\ref{width_plots_1} except using flux tubes obtained with the 
second normalization, which forces the flux tube and corresponding loop to 
be similar over the entire region of overlap.  The set of field lines 
defining the flux tube is different from that in Figure~\ref{width_plots_1} 
because the starting cross section is smaller (see Section 3.4).}
         \label{width_plots_2}
\end{figure*}

\begin{figure*}    
   \centering
  \hspace{0cm}
\includegraphics[bb= 70 200 560 650, width=16.cm ]{f11.eps}
      \caption{The flux tube shown has a non-expanding (though variable)
width when observed from the top. In the side view the expansion
is evident. This example corresponds to one of the flux tubes plotted
in Figure~\ref{width_plots_1} panel (d) (dash-dot line, ellipse at 45 degrees).}
         \label{example}
\end{figure*}

\begin{figure*}   
   \centering
  \hspace{0cm}
\includegraphics[bb= 70 70 420 800, height=19.cm]{f12.eps}
      \caption{$\Gamma^{tube}$ vs $\Gamma^{loop}$ for the
combinations $m/s$, $m/e$ and $e/s$ respectively (see 
Section~\ref{comparison} for details). Asterisks correspond to cases in 
which the loop footpoint and the flux tube footpoint are separated by less
than 10 Mm. In each case, the dashed line has 
a slope of 1, the solid line has a slope equal to the mean of 
$R$ ($=\Gamma^{loop}/\Gamma^{tube}$), and the dotted lines have slopes equal 
to the mean of $R$ plus and minus one standard deviation.
}
         \label{gammas}
\end{figure*}

\begin{figure*}   
   \centering
  \hspace{0cm}
\includegraphics[bb= 70 90 530 780, height=19.cm]{f13.eps}
      \caption{Loop width inferred from three different intensity
profile properties: the standard deviation (continuous line), the 
FWHM (dotted), and the so called
Equivalent Width (dashed) (see Section~\ref{obs_loops} and Appendix). 
The examples correspond to the loops in Figures~\ref{stdev}, 
\ref{width_plots_1}, and~\ref{width_plots_2}, panels c and d. 
The three methods give similar width values and fluctuations.
}
         \label{stdev_test}
\end{figure*}

\begin{figure*}   
   \centering
  \hspace{0cm}
\includegraphics[bb= 70 100 530 760, height=19.cm]{f14.eps}
      \caption{Scatter plots of the standard deviation of the
intensity profile vs FWHM (top) and the Equivalent Width (bottom)
for the four example loops shown in Figures~\ref{stdev}, 
\ref{width_plots_1}, and~\ref{width_plots_2} (see Section~\ref{obs_loops}
and Appendix). Units are instrument pixels. The lines correspond to 
least square fits of the data (the slopes are given in the plots).  
}
         \label{linear_fit}
\end{figure*}

\end{document}